# Cookies – Invading Our Privacy for Marketing, Advertising and Security Issues

An Analysis with interpretation

**SOWMYAN JEGATHEESAN**

**Abstract**

Privacy has been a major concern for everybody over the internet. Governments across the globe have given their views on how the internet space can be managed effectively so that there is some control on the flow of confidential information and privacy to users and as well as to data is achieved. Taking advantage of the lack of one unified body that could govern the online space with its strict and stringent rules certain websites use the text files called cookies in collecting data from users and using them for marketing them in advertisements networks and third party websites with the help of JavaScript and flash technologies. Even before many of us think what is happening around us in the online usage we are being invaded by the cookies and are targeted with their specific information that could tempt us to buy a product or service to which we were longing for in the recent past. Though it may be argued it's a kind of marketing strategy to give the customer what he wants but it can no way be something like the user is just being targeted because he has already shown interest in something and he should be disturbed until he gets hold of something. It's purely a security breach as most of the websites don't ask permission for the usage of cookies and setting them into the user's browser and the most important thing is no privacy statement is issued that the information is used for targeted marketing. This analysis paper has views from different sources, an example of such an activity that is a potential security breach and various other information of what these sites do to use the deadly cookies for commercial tricks.

# Introduction

With reference to the Katz Vs US [1] the Supreme Court has said there should be a "reasonable assumption of privacy" test and everyone online would love to have privacy and this expectation is justified as reasonable. Cookies are very important for ecommerce companies and advertising agencies. There is something called Behavioral targeting [2] that uses information linked through cookies to create a profile of a user. The user is matched with a broad profile, and they are sent adverts which are likely to interest them but this has very serious implications on the individual's opinion about online presence, privacy and personal data security. Privacy is the most important concern, even though cookies cannot harm the computer like viruses [3] and they are just text files that store data from the web servers. They just help in tracking the web usage about the particular website from a particular browser in a computer. Cookies cannot be even same for different browsers in a same computer. Most of the websites don't give the option to select whether to accept or decline cookies. Some websites don't allow the users to visit the website or perform an action just because they have rejected the usage of cookies even they very well know this could affect their business prospects. So the real question we should rise here is what is the website interested to get from the user, is it the users online usage profile or business the user gives to the website ? We seem most of the websites that don't comply with the guidelines just want to know the user much more than they need some business out of them. Many people are still ignorant about these sorts of activities by the clever websites that just gets enough information about a user and then start their marketing with the help of cookies and showing only what the website thinks the user is in need of. This is a clear breach of privacy and data security [4].

# Concerns about cookies

Cookies can be blocked as said earlier but some websites become not as effective when we don't accept them. Some of us simply to use the websites and to get the full functionality accept the cookies unwillingly and are concerned about our privacy and tracking. Using adware programs can somehow be helpful for this situation. We can also delete them after we close the browser. The websites are designed and operated in such a way that blocking the cookies completely won't be a pleasant experience for the users. In addition to the options mentioned above, a person can browse through an anonymizer [5], which will mask his or her identity. Cookies have become almost essential when browsing websites, but we should all be concerned about this particular technology as a compromise to the privacy of the individual using it.

**How this happens**

Cookies are set by the websites we visit or even sometimes by the third party websites that involve in marketing, ad serving. There are even chances that we get cookies from websites we have never heard of or never visited but those websites are associated with the websites we visit. These websites do various activities like advertizing, help in marketing the brand or product, increase revenue or even track the users. These websites may set cookies that can track the movement of users from one website to another website and can consolidate the personal data and sometimes making it available for anyone with commercial interest. The most important invasion of personal privacy is best described by this real life illustration

"Imagine, being stopped at the entrance to your local grocery store or gas station and asked to show identification in order to get in. While you're there, imagine being observed by a security guard who takes notes on what you're doing or being recorded by a hidden camera as you walk around. Information gathering cookies are not nearly so obvious, but this is essentially what they do once they get inside your computer."[6]

There is one more addition that can be added to the above situation, the websites don't simply stop at tracking, they start showing adverts, campaigns are done around you to force you to buy or look at something for the fact that you had once or frequently interested in something. This eliminates the ability for us to look for new things, in simple ways we are restricted from being shown something new and we are always shown the same things, given preferences of something we did already.

**Security Issues with Cookies**

There are security implications with cookies, some websites use cookies for providing access control schemes. A website that has a login for users may set cookies into the browser with the login credentials or even the session details; this may have issues when it is a computer used by multiple users. If this is not implemented carefully this type of system may be vulnerable to exploitation by unscrupulous third parties. It is a fact that a packet sniffer program could get in middle of the cookies when the cookies are transferred from the browser to the server and can get access to the concerned website that gave the cookies. Since DNS is used to determine the cookies that are with a particular server it may be possible to cheat or play ahead with the browser to send a cookies to a server by subverting the DNS temporarily. Compromise of our login is a greatest privacy violation that we can imagine of and it's a serious security implication as well. Some OLTP (Online Transaction Processing Systems) that use cookies should be very careful in providing privacy and security to the users of the system, the above scenario can happen and the users would never tolerate the compromise in the security. Some unauthorized transactions that may happen might have potential financial damages and customer dissatisfaction as well.

**Invasion of privacy**

There is a fear or insecurity in everyone's mind that they may be noticed by Government agencies, the user data, and communication may be intercepted in the interest of national security or any other concern similar to the Phone tapping case [7]. But many of us are unaware that there are other ways around to just peep through our browser windows and gather essential data on what we do, what we spend most time upon or what we like. We want no one to know these details or at least we think we should be requested for permission when someone is doing this for any commercial or any interest the collected data may be used. Based on assumptions, "Invasion of Privacy can be best defined as the act in which a user activity, interest, behavior, data of personal [8] nature is recorded, noticed or even tracked for any period of time without consent or with limited consent that does come up into the above category". Saying this there cannot be any violation of how the data is used after it was collected with user consent. Similar cases have been reported in Online Job seeking websites [9]. Where usage of this type of data for marketing purposes with commercial interest after they were sold to a third party. A cookie can help the websites to have good sales or generate better business, but it is not very pleasant for the user who gives business to them due to the privacy issues.

There is tremendous growth in e-commerce it would still grow at a larger pace if consumer's had a satisfying sense of security and privacy. It is the right of the users to know how the information collected is used and how securely it is handled. The current usages of cookies tell us why the consumers must be worried about their privacy violation. As mentioned before, companies use various techniques to gather information and then combine with additional information and they profile the users for behavioral targeting. Author of "Four Ethical Issues of the Information Age," talks about how profiling violates privacy when he states:

"You or I may have contributed information about us freely to each of the separate databases but that by itself does not amount to giving consent to someone to merge the data.[10] "

**Prevention of data invasion**

Most of the people who use the internet are youngsters of the information technology era. But still many people who are using the internet are unaware of the cookie technology and the serious implications behind their operation. There should be education for everyone regarding this. A survey conducted by CNN showed that the majority of the public, about 56%, did not know that websites and advertisers have the capability to track their activities by placing cookies on their hard drives [11]. This would always add to the advantage of the websites that would use the data and privacy breach to their advantage.

The "Electronic Bill of Rights" [12] By Al Gore the former Vice President of USA, was took up by him because he was concerned about the lack of privacy on the Internet. It would be a great initial step towards an electronic bill of rights through our suggestion for a possible future privacy policy. That would surely help us secure the internet which is lacking ethical and legal remedies on a wide range of issues concerning security and privacy. There should be a consensus on what has to be done to prevent it. Then we would have a solution to keep the worries about privacy as it's getting a bigger stage and it is the correct time to start and make awareness. One of the primary requirements of data protection law [13] is to ensure people are aware of how information collected about them will be used. The usage of data collected online has to be effectively reflected in the privacy policy. It would be best practice and it would enable transparency, user confidence and reputation as well.

There can be privacy advantages in third party advertising [14], in that third-party will not exactly know the real person or would not have a better identity but the content that is of personal nature when it is taken out of the first party website, even if the users are intimated about the safe use, have to be ensured the fairness it deserves and it is very essential to intimate the users about the potential involvement of many stages, agencies in the data processing there by a very transparent environment has to be created.

**Marketing choices**

The websites and companies that do the online marketing through cookies should have transparent policies and should make clear options to people about how their personal information is stored, used or transmitted. This will increase the informed choice for the users about their consent for any such activity, proper usage of the internet space freely. Being open may also help them to understand and accept the analysis of information that underpins their online activity, and which may support the services they rely on.

Providing clear opt-outs also allows people to exercise control when using web services online. Some individuals may want to visit a website without any record of their online behavior being retained. Therefore it is good practice to provide a simple means of disabling the targeting of advertising using behavioral data. It is a legal requirement under the Privacy and Electronic Communications Regulations (PECR)[15] to tell the individual when information is to be stored on their equipment, for example in the form of a conventional cookie, a Local Shared Object or flash cookie, and to give them the opportunity to refuse this. It is good idea to give users relevant advice about how they can use their web browser settings, or the choices offered on the website itself, to carry over choice over the extent to which they preserve their online anonymity, for example by ensuring that information identifying them is erased at the end of a session or any other similar remedy.

The clear idea conveyed through this paper is that users should be made aware of the way how their online presence is secure, how the information is used and the transparency levels to be maintained and the current state of operation of the cookie and online marketing techniques that is seen as a distress and provides a lot of insecurity to the users.